\documentclass[aip, apl, amsmath,amssymb, reprint,]{revtex4-2}

\usepackage[english]{babel}
\usepackage{graphicx}
\usepackage{dcolumn}
\usepackage{bm}
\usepackage{textcomp}
\usepackage[utf8]{inputenc}
\usepackage[T1]{fontenc}
\usepackage{mathptmx}
\usepackage[usenames,dvipsnames]{color}
\usepackage{booktabs}

\preprint{AIP/123-QED}
\newcommand{\ie}{i.e.,\ }

\begin{document}

\title{Large strain contribution to the laser-driven magnetization response of magnetostrictive TbFe$\mathrm{_2}$ }

\author{C.~Walz}
\affiliation{Institut f\"ur Physik \& Astronomie,  Universit\"at Potsdam,  14476 Potsdam, Germany}

\author{F.-C.~Weber}
\affiliation{Institut f\"ur Physik \& Astronomie,  Universit\"at Potsdam,  14476 Potsdam, Germany}

\author{S.-P.~Zeuschner}
\affiliation{Institut f\"ur Physik \& Astronomie,  Universit\"at Potsdam,  14476 Potsdam, Germany}

\author{K. Dumesnil}
\affiliation{Institut Jean Lamour (UMR CNRS 7198), Universit\'e Lorraine,  54000 Nancy,   France}

\author{A.~von~Reppert}
\email{reppert@uni-potsdam.de}
\affiliation{Institut f\"ur Physik \& Astronomie,  Universit\"at Potsdam,  14476 Potsdam, Germany}

\author{M.~Bargheer}
\email{bargheer@uni-potsdam.de}
\affiliation{Institut f\"ur Physik \& Astronomie,  Universit\"at Potsdam,  14476 Potsdam, Germany}
\affiliation{Helmholtz Zentrum Berlin,  12489 Berlin, Germany}

\date{\today}

\begin{abstract}
We investigate strain-induced contributions to the transient polar magneto-optical Kerr effect response in laser-excited terfenol. The tr-MOKE signals obtained from $\mathrm{TbFe_2}$ films with and without glass capping exhibit distinct signatures associated with transient strain. We experimentally observe the arrival of strain pulses via the reflectivity change. The tr-MOKE response measured without changing the pump-probe geometry is  delayed by several picoseconds. This suggests a genuine magnetization response as opposed to instantaneous changes of optical constants as the origin of the signal. We model the propagation of longitudinal acoustic picosecond strain pulses and incorporate the inverse magnetostriction effect via a magnetoelastic term in the effective field of the Landau–Lifshitz–Gilbert equation with large damping. This reproduces not only the delay of the pulsed response, but also unveils the dominant contribution of quasi-static strain to the magnetization dynamics due to the thermal expansion in the optically probed near-surface region. 
Our experiments exemplify that purely longitudinal strain along the out-of-plane direction of the thin film enables efficient magnetoelastic coupling  via the shear strain components arising in the oblique crystallographic frame of reference.
\end{abstract}

\maketitle
\begin{onecolumngrid}
\begingroup\footnotesize
\noindent\rule{\textwidth}{0.4pt}\par
\textbf{Credits:} “This article may be downloaded for personal use only. Any other use requires prior permission of the author and AIP Publishing.
This article appeared in C. Walz, F.-C. Weber, S.-P. Zeuschner, K. Dumesnil, A. von Reppert, M. Bargheer; \textit{Large strain contribution to the laser-driven magnetization response of magnetostrictive} TbFe\textsubscript{2}. \textit{Appl. Phys. Lett.} \textbf{127}, 052406 (2025), and may be found at \href{https://doi.org/10.1063/5.0279959}{https://doi.org/10.1063/5.0279959}. \\
\textbf{Copyright:} © 2025 Author(s). CC BY 4.0. \href{https://creativecommons.org/licenses/by/4.0/}{https://creativecommons.org/licenses/by/4.0/}
\par\noindent\rule{\textwidth}{0.4pt}
\endgroup
\end{onecolumngrid}
\twocolumngrid
The fascinating ultrafast magnetization dynamics of thin films following from femtosecond lightpulse absorption\cite{beau1996} is inherently accompanied by a lattice strain that arises both as coherent picosecond strain pulses and as quasi-static thermal expansion \cite{thom1986}. Magnetoelastic coupling effects have initially been observed as strain-driven magnetization changes via all-optical pump-probe techniques\cite{sche2010, kim2012,yang2021b} but ultrafast diffraction techniques also identified strain dynamics induced by magnetization changes\cite{rett2016, repp2020, reid2018,  matt2023a}. The underlying phenomenon of magnetostriction is common to all magnetically ordered materials, and its application potential has increased with the discovery of rare earth iron alloys exhibiting room-temperature giant magnetostriction in the 1970s\cite{engd2000}. Terfenol (TbFe$_\mathrm{2}$) is hitherto recognized for exhibiting the largest room temperature magnetostriction effect, with relative lattice constant changes exceeding $10^{-3}$ at magnetic saturation\cite{clar1972}. Multiple variants of its related low-anisotropy composite Tb$_{0.3}$Dy$_{0.7}$Fe$_2$ (Terfenol-D) are established in applications such as electronically driven magnetostrictive ultrasound transducers and sensors for low-frequency applications\cite{pati2024,engd2000}. However, its femtosecond laser-driven magnetization dynamics are less explored despite the prospect of large magnetoelastic coupling effects that may even yield magnetization switching effects via high-amplitude picosecond strain pulses\cite{kova2013}.

Ultrafast magnetoelastic effects have been initially discovered via picosecond strain pulse driven magnetization precession effects\cite{sche2010, kim2012}, that can  even drive higher-order standing spinwaves \cite{bomb2012, deb2021b}.  Such large-amplitude picosecond strain pulses were recently found to be capable of altering not only the orientation but also the magnitude of the magnetization\cite{pfaf2025} and may emit coherent magnons alongside their propagation in a magnetic material\cite{fila2025}. 
A growing number of studies suggest that also quasi-static strain effects may contribute significantly to the laser induced magnetization response even in materials with relatively weak magnetostriction, such as Ni, NiFe alloys and  Co\cite{shin2022, shin2023}. Nevertheless, extracting the magnetoelastic contributions from the overall ultrafast magnetization dynamics remains challenging due to concurrence of demagnetization, and time-dependent changes in magneto-crystalline anisotropy\cite{linn2011a}. Related experimental approaches encompass double-pulse excitation schemes\cite{matt2024}, variations of the magnetoelastic coupling strength via material-composition\cite{shin2023}, analysis of spectral shifts of the magneto-optical response\cite{shin2025} or dedicated heterostructure design that tailors the strain pulses and thermal energy flow \cite{jare2024}. 

Previous investigations using time-resolved magneto-optical Kerr effect (trMOKE) measurements on TbFe$_2$ films have indeed shown distinct features associated with coherent strain pulses\cite{zeus2019,parp2021}. However, ambiguity persists regarding whether these strain-induced signatures reflect actual magnetization vector alterations or merely photo-elastic changes in magneto-optical constants\cite{thev2010}.

In this work, we present a detailed comparison between trMOKE and reflectivity measurements on TbFe$_2$, providing evidence that the observed strain signatures result from  magnetization tilting. We identify a distinct picosecond delay between the arrival of picosecond strain pulses, that is detected via reflectivity, and the onset of the trMOKE response, consistent with predictions from a strongly damped macrospin evolution based on the LLG equation. 
Additionally, the analysis of multiple delayed strain pulse echoes in glass-capped samples allows temporal separation and precise calibration of the magnetization response to strain pulses. Extending this calibration, we demonstrate that quasi-static magnetoelastic effects constitute the major contribution to the total observed trMOKE signal in TbFe$_2$ especially at pump-probe delays beyond 100ps.

\begin{figure}[htbp]
\centering
\includegraphics[width = \columnwidth]{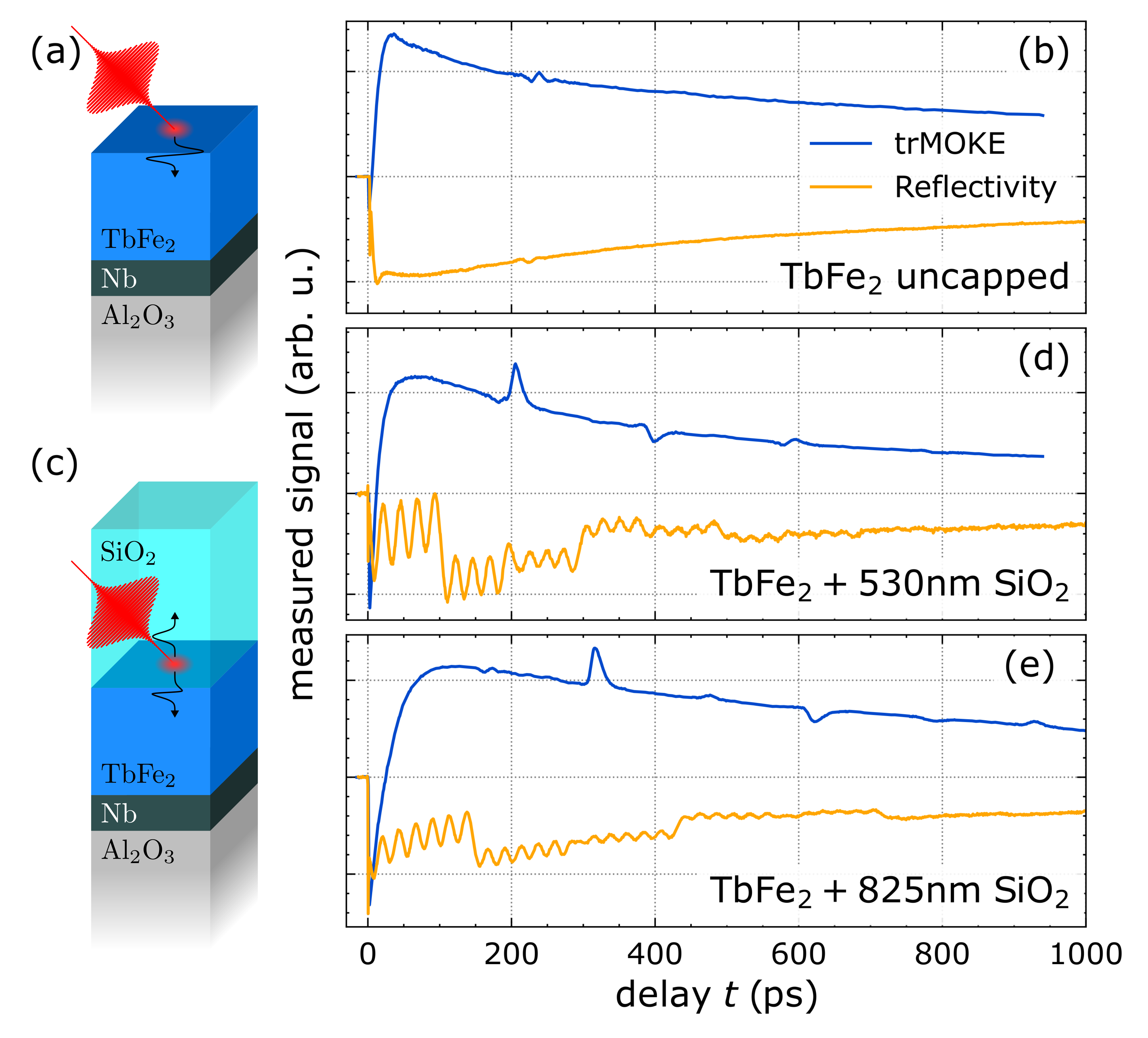}
  \caption[representative trMOKE]{\textbf{ Representative trMOKE responses and reflectivity change in (110)-oriented $\mathrm{TbFe_2}$ films:} (a) Schematic of an uncapped $\mathrm{TbFe_2}$ sample structure and (b) corresponding measured transient signals. (c) Schematic of $\mathrm{TbFe_2}$  films capped with $\mathrm{SiO_2}$  layers of varying thicknesses and corresponding transient signals measured for (d) 530\,nm and (e) 825\,nm $\mathrm{SiO_2}$ capping layers. The presence and characteristics of picosecond strain pulses can be clearly distinguished in both the trMOKE (magnetic) and reflectivity $\Delta R$ (primarily non-magnetic) responses, highlighting the impact of capping layers on strain dynamics and magnetization response.  The oscillations observed in the transient reflectivity of the capped samples (d, e) arise from Brillouin scattering of light from propagating picosecond strain pulses within the transparent glass capping.\label{fig:Fig1}}
\end{figure}

The samples were grown via molecular beam epitaxy following the procedures outlined in previous works\cite{moug1999,moug2000,parp2021} and share a common structure: (110)-oriented TbFe$_\mathrm{2}$ films are deposited on a 50\,nm niobium (Nb) buffer layer on sapphire (Al$_2$O$_3$) substrates. One sample is capped with a thin 2\,nm titanium (Ti) layer (Fig. \ref{fig:Fig1}(a)), while the other two are capped with 530\,nm and 820\,nm thick fused silica (SiO$_\mathrm{2}$) layers, respectively (Fig. \ref{fig:Fig1}(c)). The TbFe$_\mathrm{2}$ film thickness is approximately 450\,nm for the Ti-capped sample (referred to as "uncapped") and around 350\,nm for the glass-capped samples. The unintended variation in TbFe$_2$ thickness between the samples does not impact the interpretation of our results, since both samples are significantly thicker than the optical penetration depth that is below 20\,nm for both pump and probe pulses. Consequently, the thickness differences primarily influence only the arrival timing of the picosecond strain pulse reflections within the sample structure.

The magnetization dynamics are induced by femtosecond laser excitation and the $m_z$-component is monitored via a polar trMOKE setup. A mode-locked, amplified Ti:sapphire laser produces linearly polarized 150\,fs pump and probe pulses. The pump beam, operating at a fundamental wavelength of 800\,nm, excites the sample, while the probe beam frequency-doubled to 400\,nm monitors the response. This two-color scheme eliminates optical interference between the pump and probe beams and reduces background contributions from the pump light, which can be effectively filtered before reaching the detector. The reflected probe polarization rotation is analyzed using a Wollaston prism that separates the orthogonal polarization components onto a pair of balanced photodiodes. An electromagnet applies a variable B-field of up to 1.3\,T perpendicular to the sample surface. For each measurement, signals recorded under opposite magnetic field directions are subtracted, effectively eliminating magnetic field-independent contributions from the trMOKE signal. A detailed schematic of the trMOKE setup and additional experimental details are provided in Fig.~S1 of the Supplementary Material and further described in Refs.\cite{will2019a,deb2022}.

Transient reflectivity measurements were performed under identical laser-excitation conditions and at the same probe spot as the magneto-optical (tr-MOKE) experiments. Crucially, our approach ensures that the detection of strain pulses via transient reflectivity changes occurs  in immediate succession  and spatially overlapped with the magnetization dynamics, as the detection pathway is altered only after the probe beam has interacted with the sample. Figure~\ref{fig:Fig1} shows representative tr-MOKE signals, recorded at an external field of $1.3\,\mathrm{T}$ and an excitation fluence of $8\,\mathrm{mJ/cm^2}$, alongside corresponding reflectivity changes for all three sample structures.

We begin the discussion and analysis of our experimental results with the trMOKE response of the basic uncapped TbFe$_2$ structure shown in Fig.~1(b). Immediately after pump laser excitation at $t=0$, the trMOKE signal exhibits a rapid, sub-picosecond drop followed by a fluence-dependent rise that exceeds the initial level. Over several hundreds of  picoseconds, the signal gradually relaxes back to the pre-excitation level. This complex evolution depends both on the excitation fluence and the initial tilt of the magnetization relative to the out-of-plane direction, since the applied 1.3\,T field is insufficient to fully saturate the magnetization.  This also prevents a quantitative analysis of the recorded trMOKE signal, as the full hysteresis amplitude cannot be used for calibration. 
A complete interpretation of the trMOKE data, including the interplay between Fe and Tb sublattice demagnetization, and the out-of-plane tilting of the overall magnetization induced by a reduction in magnetocrystalline anisotropy, is beyond the scope of this letter.

The primary focus of our study is the rapid modulation observed in both the trMOKE and transient reflectivity signals at approximately 230\,ps.  The timing of this picosecond strain pulse signature  is independent of the external magnetic field amplitude and laser excitation fluence, as demonstrated by additional measurements (see Fig.~S4  in the Supplementary Material). It marks the arrival of reflected strain pulses in the optically probed near-surface region of the TbFe$_2$ film, consistent with previous ultrafast X-ray diffraction findings\cite{zeus2019}.

To accurately capture the shape and timing of the strain pulses observed in the trMOKE and reflectivity signals, we simulated the laser-induced strain response by numerically solving the elastic wave equation using the \textsc{udkm1Dsim} toolbox\cite{schi2021}. This simulation approach assumes a one-dimensional linear chain model and a one-temperature model to describe the laser-induced lattice dynamics. It has previously been demonstrated to effectively reproduce the strain response in Terfenol thin film structures probed by UXRD\cite{zeus2019, matt2023b}. The relevant thermophysical parameters for our samples are summarized in Sec. 2 of the Supplementary Material. Figure \ref{fig:Fig2}(a) presents the simulated evolution of the strain field $\eta_{zz}(z,t)$ along the out-of-plane direction $z$ as a function of delay $t$ after laser excitation, along with a schematic representation of the 400\,nm probe pulse's optical penetration profile, which indicates an $18\,\text{nm}$ penetration depth that is common to all three sample structures (see Fig. S2 in the Supplementary Material).

The strain simulation demonstrates that after laser excitation the pumped near-surface region undergoes thermal expansion, generating a quasi-static strain. Concurrently, a coherent strain pulse is launched, propagating through the sample where it is partially reflected and transmitted at the interfaces due to the acoustic impedance mismatch between layers. In agreement with the transient reflectivity and trMOKE data, the first echo of this strain pulse reaches the probed region at approximately 230\,ps, resulting in a modulation of both the magneto-optical and reflectivity signals (Fig. \ref{fig:Fig2}(b)). 
Although the reflectivity does not provide a quantitative measure of the strain amplitude, the simulation matches the strain values previously measured via UXRD on the exact same sample \cite{zeus2019}.

\begin{figure}[htbp]
\centering
\includegraphics[width = \columnwidth]{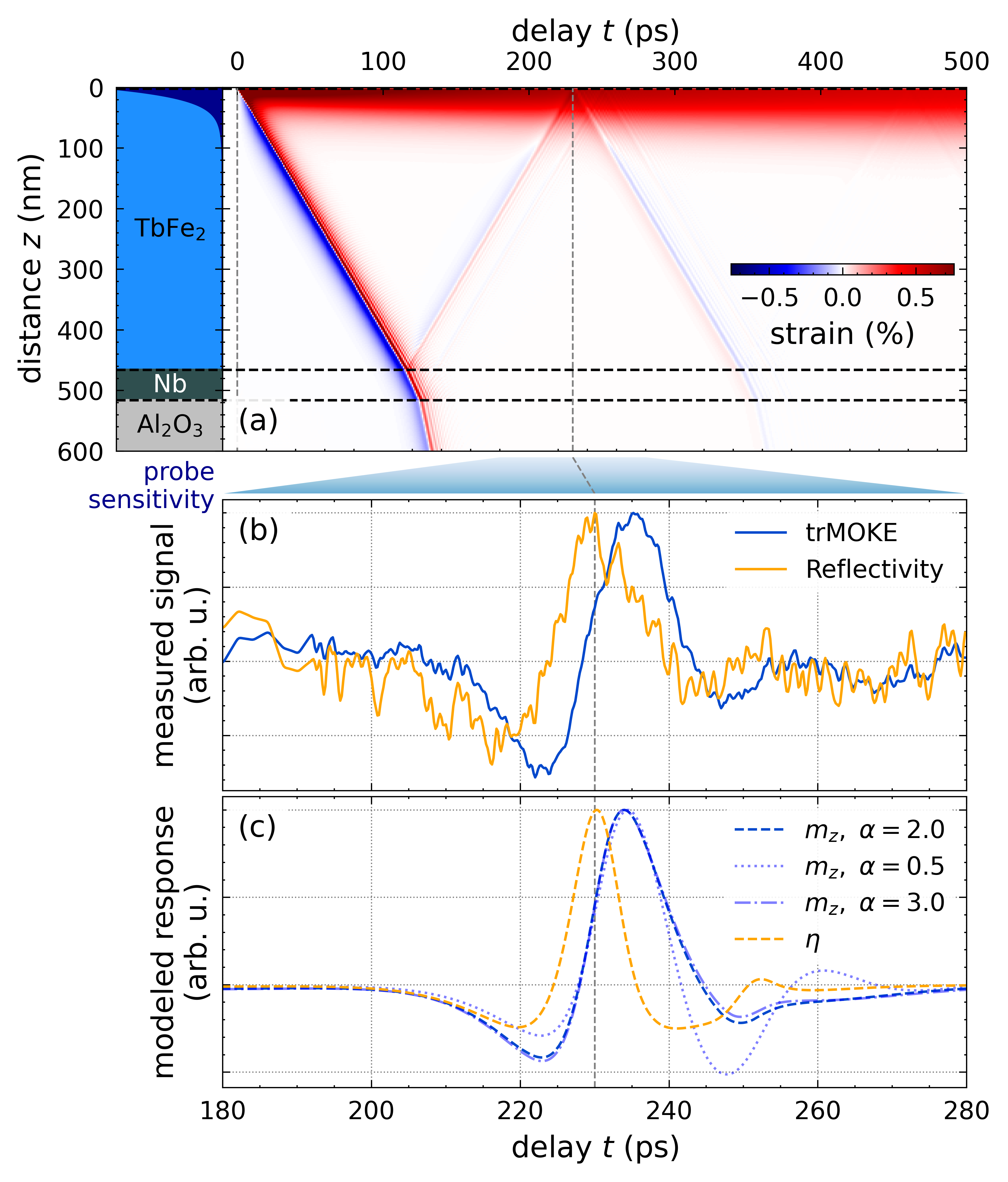}
  \caption[Analysis of the trMOKE response in uncapped TbFe2:]{\textbf{Analysis of the trMOKE response in uncapped $\bm{\mathrm{TbFe_2}}$:} (a) Simulated time- and space-resolved strain profile illustrating the propagation of picosecond strain pulses within the multilayer sample structure depicted on the left. (b) Measured background-subtracted trMOKE and reflectivity responses highlighting magnetic and non-magnetic contributions, respectively. (c) Modeled temporal evolution of the magnetization component ($m\mathrm{_z}$), driven by the strain ($\eta_{zz}$), whose peak precedes the magnetization response by approximately 5\,ps, consistent with the experimental observations shown in (b). Both signals are background-subtracted to extract the strain signature only. {A range of modeled $m_\mathrm{z}(t)$ curves illustrate the effect of varying the damping parameter $\alpha$.} \label{fig:Fig2}}
\end{figure}

Figure 2(b) presents a zoomed view of the trMOKE data alongside the time-resolved reflectivity changes of the uncapped TbFe$_\mathrm{2}$ sample at 230 ps, when the reflection from the TbFe$_\mathrm{2}$/Nb interface reaches the surface. The slowly varying background is subtracted from both signals. The delay of the trMOKE response by approximately 5\,ps relative to the reflectivity change is an experimental fact, since both signals were pumped and probed under identical laser-excitation conditions. The delay excludes a modulation of the optical constants as the origin of the strain-driven trMOKE signal because that mechanism would produce an instantaneous response. 

Instead, the observed signal shape and delay can be rationalized as a strongly damped magnetization precession driven via a magnetoelastic coupling from picosecond strain pulses. To capture this behavior, we use a %simplified
Landau–Lifshitz–Gilbert (LLG) model for a single macrospin representing the average magnetization in the optically probed region. In accordance with previous phenomenological models \cite{linn2011a,jare2024, yang2021b} the LLG equation reads:

\begin{equation}
\frac{\partial\vec{m}}{\partial t}=-\frac{\mu_{0}\gamma}{1+\alpha^{2}}\vec{m}\times\vec{H}_{\text{eff}}-\frac{\mu_{0}\gamma\alpha}{1+\alpha^{2}}\vec{m}\times\left(\vec{m}\times\vec{H}_{\text{eff}}\right),
\end{equation}
where $\vec{m}=\vec{M}/M_{S}$ is the normalized magnetization vector, $\mu_{0}$ the magnetic permeability, $\gamma$ the gyromagnetic ratio and the Gilbert damping constant $\alpha$. Our interpretation relies on a large damping $\alpha=2$, which is expected in materials with strong spin-orbit interaction\cite{he2013}. This high damping suppresses any subsequent magnetization oscillations that are reported in other prominent magnetoacoustic studies (e.g., GaMnAs \cite{sche2010}, Ni\cite{kim2012} and Bi:YIG\cite{deb2018}).

In our model, the effective field $\vec{H}_\text{eff}$ contains an external field $\vec{H}_{\text{ext}}$, a demagnetization field $\vec{H}_{\text{D}}$, a cubic magneto-crystalline anisotropy field $\vec{\tilde{H}}_{\text{ani}}$ and a magnetoelastic field $\vec{\tilde{H}}_{\text{me}}$:

{\small
\begin{subequations}
\begin{align}
\vec{H}_\mathrm{D} &= -M_{S}\,m_{z}\,\hat{z},\\[1ex]
\vec{\tilde{H}}_{\text{ani}} & =-\frac{2K_{1}}{\mu_{0}M_{S}}\begin{pmatrix}
\tilde{m}_{x}\left(\tilde{m}_{y}^{2}+\tilde{m}_{z}^{2}\right)\\
\tilde{m}_{y}\left(\tilde{m}_{x}^{2}+\tilde{m}_{z}^{2}\right)\\
\tilde{m}_{z}\left(\tilde{m}_{x}^{2}+\tilde{m}_{y}^{2}\right)
\end{pmatrix}-\frac{2K_{2}}{\mu_{0}M_{S}}\begin{pmatrix}
\tilde{m}_{x}\tilde{m}_{y}^{2}\tilde{m}_{z}^{2}\\
\tilde{m}_{y}\tilde{m}_{x}^{2}\tilde{m}_{z}^{2}\\
\tilde{m}_{z}\tilde{m}_{x}^{2}\tilde{m}_{y}^{2}
\end{pmatrix},\\
\vec{\tilde{H}}_{\text{me}} & =-\frac{2b_{1}}{\mu_{0}M_{S}}\begin{pmatrix}
\tilde{\eta}_{xx}\tilde{m}_{x}\\
\tilde{\eta}_{yy}\tilde{m}_{y}\\
\tilde{\eta}_{zz}\tilde{m}_{z}
\end{pmatrix}-\frac{b_{2}}{\mu_{0}M_{S}}\begin{pmatrix}
\tilde{\eta}_{xy}\tilde{m}_{y}+\tilde{\eta}_{xz}\tilde{m}_{z}\\
\tilde{\eta}_{xy}\tilde{m}_{x}+\tilde{\eta}_{yz}\tilde{m}_{z}\\
\tilde{\eta}_{xz}\tilde{m}_{x}+\tilde{\eta}_{yz}\tilde{m}_{y}
\end{pmatrix}.
\end{align}
\end{subequations}
}

Here, $K_{1,2}$ represent the cubic magneto-crystalline anisotropy constants, while $b_{1,2}$ denote the magnetoelastic coupling parameters, with $|b_{2}|\gg |b_{1}|$ for TbFe$_\mathrm{2}$ \cite{parp2021}. Quantities defined in the crystallographic reference frame are indicated by a tilde. 
The anisotropy field $\vec{\tilde{H}}_{\mathrm{ani}}$ and magnetoelastic field $\vec{\tilde{H}}_{\mathrm{me}}$ are rotated into the thin‐film sample frame of reference for the (110)-orientation as detailed in Sec.~S.3 of the Supplementary Material. To compute the magnetoelastic field in response to the purely longitudinal, out‐of‐plane strain $\eta_{zz}(t)$, we first transform $\eta_{zz}(t)$ into crystallographic reference frame, yielding normal ($\tilde\eta_{ii}$) and shear ($\tilde\eta_{ij}$) strain components of equal magnitude. Since $|b_2| \gg |b_1|$, only the shear terms contribute significantly. Transforming back gives:
\begin{equation}
\vec{H}_{\text{me}}\approx-\frac{1}{\sqrt{2}}b_{2}\eta_{zz}\left(\begin{array}{c}
-m_{x}\\
0\\
m_{z}
\end{array}\right).
\end{equation}
Shear thus plays a twofold role for the observed strain-driven magnetization response: \ie shear components of the elastic tensor $C_{ijkl}$ affect strain pulse arrival time via the longitudinal sound velocity,
$v_{[110]} \propto \sqrt{C_{1111} + C_{1122} + 2\,C_{1212}}$ and, more importantly, shear strain components ($\tilde\eta_{ij}$) predominantly drive the magnetoelastic response via the $b_2$ coupling term.

To accurately represent the time-dependent strain within the optically probed region $\eta_{zz}(t)$, we apply a weighting to the simulated strain distribution $\eta_{zz}(z,t)$ (Fig.~2(a)), using the optical penetration profile of the 400\,nm probe, characterized by a penetration depth of approximately 18\,nm  in accordance  with a prior spectroscopic ellipsometry analysis \cite{zeus2019}.  This weighted strain alters the effective magnetocrystalline anisotropy constants, represented as a time- and strain-dependent magnetoelastic field in our LLG model. Its rapid modulation during the arrival of the strain pulse acts as a transient torque, directly inducing the strongly damped precession displayed in Fig.~2(c). The modeled $m_\mathrm{z}$ response qualitatively reproduces both the observed trMOKE signal shape and the 5\,ps delay between the strain pulse arrival and the onset of magnetization precession.
We demonstrate the effect of varying the damping constant $\alpha$ from 0.5 to 3. For lower values ($\alpha \leq 0.5$), the magnetization exhibits precession following the strain pulse. This oscillation is progressively suppressed at higher damping, and for $\alpha \geq 2$, the dynamics become overdamped and show minimal further change with increasing $\alpha$.

\begin{figure}[htbp]
\centering
\includegraphics[width = \columnwidth]{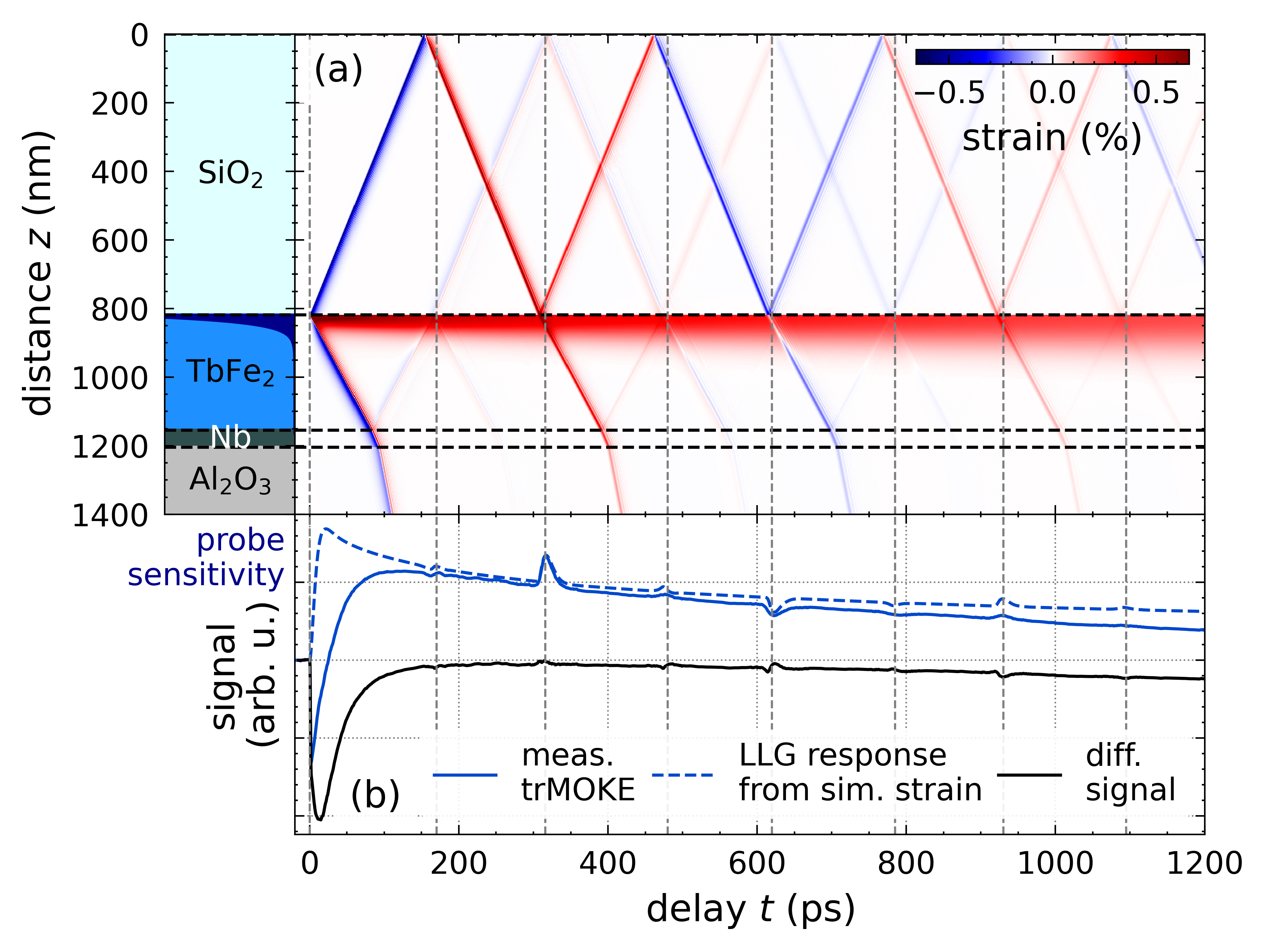}
  \caption[Analysis of the trMOKE response in capped TbFe2:]{\textbf{Analysis of the trMOKE response in glass-capped $\bm{\mathrm{TbFe_2}}$:} (a) Spatiotemporal evolution of strain induced by ultrafast excitation, simulated using the udkm1Dsim toolbox\cite{schi2021}. Clear acoustic reflections at the interfaces of the multilayer structure are visible. The probe sensitivity region within the $\mathrm{TbFe_2}$ layer is highlighted in dark blue. (b) Comparison of the measured trMOKE signal (solid blue) with the simulated magneto-optical response derived from the strain profile using the Landau-Lifshitz-Gilbert (LLG) equation (dashed blue). Dashed vertical lines indicate the arrival times of strain echoes at the $\mathrm{TbFe_2}$ surface. The nearly negligible difference signal (solid black) for delays beyond 200\,ps emphasizes that the magnetization dynamics at large delay times can be fully explained by magnetoelastic contributions arising from quasi-static strain and propagating strain pulses.}  \label{fig:Fig3}
\end{figure}

Due to good acoustic impedance matching between TbFe$_2$ and Nb, the uncapped sample exhibits only a single strain pulse echo. In contrast, the impedance mismatch between SiO$_2$ and TbFe$_2$ in the capped samples results in multiple reflections of the unipolar strain pulse that propagates within the glass layer, yielding up to three observable echoes at 310\,ps, 617\,ps, and 924\,ps (cf. Figs.~\ref{fig:Fig1}(d,e) and \ref{fig:Fig3}). Secondary echoes, arising from reflections at the TbFe$_2$/Nb interface, appear approximately 165\,ps after each primary echo (i.e., at 170\,ps, 475\,ps, and 782\,ps). However, this strain response cannot be directly probed by transient reflectivity measurements because strong Brillouin oscillations (cf. Figs.\ref{fig:Fig1}(d,e)), caused by interference between probe light reflected from the sample surface and the propagating strain pulse in the SiO$_2$, mask these signals. Thus, unlike in the uncapped sample, a direct comparison between reflectivity and trMOKE is not feasible. Instead, the multiple strain echoes observed in the trMOKE response allow us to calibrate the magnetoelastic contribution by extracting the strain response from our simulations.

Again, we calculate the weighted strain response from simulations (Fig.~\ref{fig:Fig3}(a)) to model the magnetoelastic field. The strain-driven LLG model accurately reproduces both the timing and the rounded shape of the coherent strain pulses (Fig.~\ref{fig:Fig3}(b)), as well as the slowly changing quasi-static strain contribution. 
Both strain contributions couple to the magnetization through the same magnetoelastic mechanism previously discussed for the uncapped sample.
 Since the peaks in the trMOKE response arise solely from the magnetoelastic contribution, they can be used to calibrate the $\alpha$, $K_{1,2}$ and $b_{1,2}$ parameters in the LLG.
This calibration enables us to model the magnetoelastic contribution to the trMOKE signal over the full delay range, including the quasi-static component. Consequently, the difference between the experimental trMOKE signal and the strain-driven LLG response approximates the additional intrinsic, non–strain-driven magnetization dynamics. This difference, shown as the black line in Fig.~\ref{fig:Fig3}(c), exhibits a rapid initial drop, which is indicative of ultrafast demagnetization, followed by a recovery within the first 200\,ps.

In summary, we identify a previously unrecognized delay between the arrival of strain pulses and the onset of the trMOKE response in TbFe$_2$, demonstrating that propagating picosecond strain pulses drive a strongly damped magnetization precession. This delay is well reproduced by an LLG model that incorporates the time-dependent strain within the probed sample region, which allows for a direct calibration of the magnetization response to strain. 
In addition to the signals arising from the propagating picosecond strain pulses, the same magnetoelastic coupling mediates the tr-MOKE signal associated with quasi-static strain induced by thermal expansion in the laser-excited near surface region. The  large  contribution of magnetoelastic effects to the trMOKE signals is likely a general characteristic of materials with strong magnetostriction. For our (110)-oriented films it is the shear component in the crystallographic frame of reference that predominantly couples the  longitudinal acoustic wave to magnetization precession, parametrized by $b_2$.

The sample design presented here, featuring a thick magnetic layer or a transparent capping layer that acts as an acoustic delay line, enables the temporal separation of strain-induced effects from laser-driven thermal changes of the magnetization and anisotropy field contributions. 
This approach can be readily applied to the study of magnetoelastic coupling in a broad range of material systems. Moreover, analyzing the transient reflectivity change of the probe pulses in addition to the polarization change from trMOKE offers a straightforward, yet powerful, approach to correlate strain and magnetization response within the same experiment. An explicit numerical modeling of the optically probed strain dynamics within multilayered heterostructures enables the identification of key features governing strain-driven magnetization dynamics. Ultimately, our results demonstrate that both picosecond strain pulses and quasi-static lattice expansion, are essential contributors to the trMOKE response and must be considered when analyzing laser-induced magnetization precession and switching phenomena.

\section*{Supplementary Material}

The supplementary material provides additional details that support the findings presented in the main text. It includes:  
(1) a description of the experimental setup,  
(2) thermophysical and magnetic properties of the studied samples,  
(3) implementation details, assumptions, and limitations of the LLG model, and  
(4) supplementary data showing fluence- and magnetic field-dependent trMOKE measurements.

\section*{Data Availability}
The data and simulation scripts of this study are openly available in Zenodo at \url{https://doi.org/10.5281/zenodo.15309120}.

\begin{acknowledgments}
We acknowledge the DFG for financial support Project No. 328545488—TRR 227, project A10.
\end{acknowledgments}
 
\bibliography{references.bib}
\clearpage

\appendix
\onecolumngrid

\part*{Supplementary Material}

%%%%%%%%%%%%%%%%%%%%%%%%%%%%%%%%%%%%%%%%%%%%%%%%%%%%%%%%%%%%%%%%%%%%%%%%%%%%%%
\bibliographystyle{apsrev4-2}

\newcommand{\etal}{et al.\ }
\newcommand{\iu}{{i\mkern1mu}}

\preprint{APS/123-QED}

\title{Supplementary material to: \\ Large strain contribution to the laser-driven magnetization response of magnetostrictive TbFe$\mathrm{_2}$}

\author{C.~Walz}
\affiliation{Institut f\"ur Physik \& Astronomie,  Universit\"at Potsdam,  14476 Potsdam, Germany}

\author{F.-C.~Weber}
\affiliation{Institut f\"ur Physik \& Astronomie,  Universit\"at Potsdam,  14476 Potsdam, Germany}

\author{S.-P.~Zeuschner}
\affiliation{Institut f\"ur Physik \& Astronomie,  Universit\"at Potsdam,  14476 Potsdam, Germany}

\author{K. Dumesnil}
\affiliation{Institut Jean Lamour (UMR CNRS 7198), Universit\'e Lorraine,  54000 Nancy,   France}

\author{A.~von~Reppert}
\email{reppert@uni-potsdam.de}
\affiliation{Institut f\"ur Physik \& Astronomie,  Universit\"at Potsdam,  14476 Potsdam, Germany}

\author{M.~Bargheer}
\email{bargheer@uni-potsdam.de}
\affiliation{Institut f\"ur Physik \& Astronomie,  Universit\"at Potsdam,  14476 Potsdam, Germany}
\affiliation{Helmholtz Zentrum Berlin,  12489 Berlin, Germany}
\date{\today}
\maketitle
\renewcommand{\thefigure}{S\arabic{figure}} %numbers figures with S1
\renewcommand{\thesection}{S\arabic{section}} 
\renewcommand{\thetable}{S\arabic{table}}

\section{Experimental setup}
\begin{figure}[htbp]
\centering
\includegraphics[width = .68\textwidth]{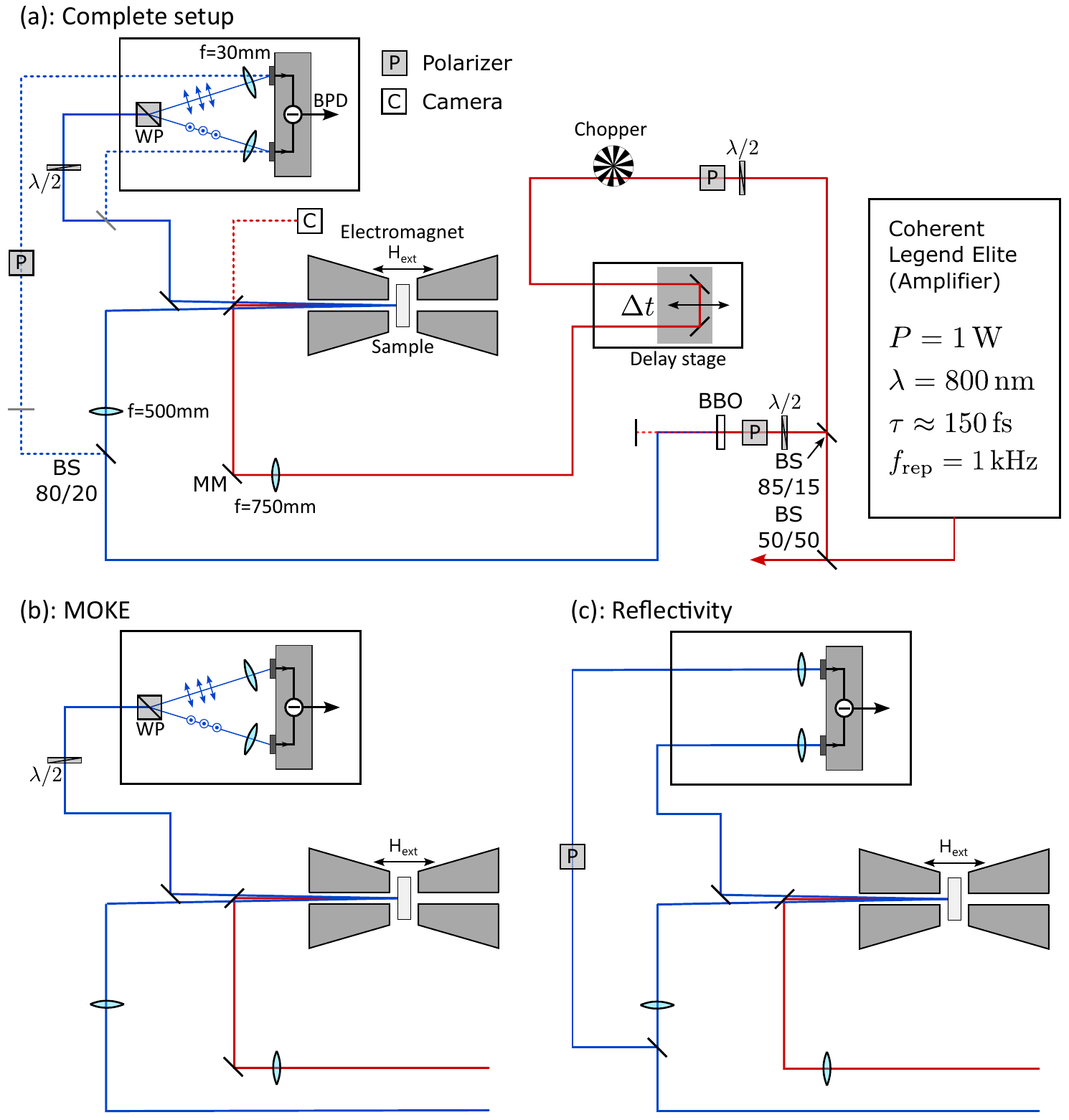}
\caption{\label{fig:schematic_setup}\textbf{Sketch of the experimental setup:} Pump-probe experiments for the detection of the time-resolved magneto-optic kerr effect and the reflectivity change upon laser-excitation with femtosecond laser pulses.\label{fig:S1}}
\end{figure}

In the reported experiments we employ a modified version of the time-resolved magneto-optics pump-probe setup that was described in previous publications\cite{will2019a, deb2022}.   It is schematically depicted in Fig.~\ref{fig:S1}.

In brief: Pump and probe pulses are generated by a 1\,kHz amplified femotosecond Laser system (Coherent Legend Elite) seeded by an 80\,MHz oscillator (Coherent Mira) pulses at a central wavelength of approximately 800\,nm. A beam splitter separates 15\,\% of the light for the probe pulses, while the remainder forms the pump. The pump-pulse intensity is controlled via a motor-controlled $\lambda$/2-waveplate and polarizer, and the beam is chopped at 500\,Hz to alternate between pumped and unpumped signals in the experiment. A motorized mechanical delay stage is used to adjust the pump–probe timing. The pump is focused onto the sample through a dichroic mirror—which reflects 800\,nm and transmits 400\,nm and its position is stabilized using a motorized mirror with camera feedback.

The probe is first intensity-adjusted using a $\lambda$/2-waveplate and polarizer, then frequency doubled from 800\,nm to 400\,nm in a Beta Barium Borate (BBO) crystal. A secondary beam splitter extracts a reference for transient reflectivity (blocked during trMOKE experiments), and the remaining beam is focused onto the sample at near-normal incidence in a polar MOKE geometry. The reflected light is directed via a third beam splitter to a detection setup, employing a Wollaston prism and balanced photodiodes, where the polarization rotation is detected as described for example by Legaré et al.\cite{lega2022}. For reflectivity measurements, a flip mirror redirects the probe before the $\lambda$/2-waveplate so that one photodiode records the reference intensity while the other detects the reflected beam. This enables a  balanced detection scheme that mitigates noise arising from  shot to shot fluctuations of the laser-intensity.

\section{Sample properties}
The following section provides additional details on the sample and material parameters used to model the strain and magnetization responses discussed in the main manuscript.

\subsection{Material parameters}
TbFe$_2$ belongs to the family of rare-earth iron (REFe$_2$) compounds and crystallizes in a cubic C15 Laves-phase structure.\cite{engd2000} Each Tb atom contributes a significant magnetic moment of approximately $9.3\,\mu_\mathrm{B}$, which couples antiferromagnetically to the magnetic moments of the two Fe atoms (each $\sim1.65\,\mu_\mathrm{B}$).\cite{busc1970} This antiferromagnetic coupling results in a ferrimagnetic configuration with a net magnetic moment of about $(9.3 - 2\times 1.65)\,\mu_\mathrm{B}\approx 6\,\mu_\mathrm{B}$ per formula unit. Magnetic hysteresis loops measured along both in-plane and out-of-plane directions have been previously reported by Parpiev \textit{et al.}\cite{parp2021}. Their findings indicate that a saturation field of approximately $2\,\mathrm{T}$ is necessary to fully align the magnetization along the out-of-plane direction. Consequently, the maximum field strength of $1.3\,\mathrm{T}$ available from our electromagnet allows only for partial out-of-plane tilting of the magnetization. The thermophysical parameters used for the strain simulations and the relevant magnetic properties required to model magnetization dynamics via the Landau–Lifshitz–Gilbert (LLG) equation are reported in this section.

To model the excited strain response in our heterostructures, we use the \textsc{udkm1Dsim} Python toolbox \cite{schi2021}. This code numerically solves the continuum elastic wave equation using a linear-chain model of masses and springs driven by a spatiotemporally varying stress generated by the laser-induced energy density. To calculate the stress profile, it solves the one-dimensional heat diffusion equation, which requires a set of thermophysical properties and refractive indices that are listed in Table~\ref{tab:Modeling_parameters}. The initial energy deposition from the pump-laser pulses is derived from the built-in optical transfer matrix model that uses the provided complex refractive indices. A detailed description of the modeling approach is provided in a recent overview article \cite{matt2023b}, and the simulation parameters used here are listed in Table~S1. Our modeling approach closely follows the work of Zeuschner et al. \cite{zeus2019}, where the obtained strain response was further validated via layer-specific ultrafast X-ray diffraction.

\begin{table}[htbp]
\centering
\caption{\label{tab:Modeling_parameters} Thermophysical parameters employed in the \textsc{udkm1Dsim} modeling of the laser-driven strain response. The sound velocity for TbFe$_2$ along the (110) direction (denoted with $^{*}$), is derived from the elastic constants reported by Patrick et al. \cite{patr2020}, and the sound velocity for SiO$_2$ (denoted with **), determined from time-domain Brillouin scattering oscillations observed in the reported reflectivity measurements, are provided. The material specific acoustic impedance ($Z$) is computed via $Z=\rho v$ to show that the main acoustic impedance mismatch occurs at the glass TbFe$_2$ interface.}
\begin{tabular}{lccccc}
\toprule 
 & TbFe$_2$ & $\text{SiO}_{2}$ & Nb & $\text{Al}_{2}\text{O}_{3}$ & Ti\tabularnewline
\midrule
\midrule 

out-of-plane

lattice constant $c\,\left(\text{\AA}\right)$
 & 

$10.42$

$(110)$ \cite{coey2013}
 & --- & 

$4.66$

$(110)$ \cite{haba2013}
 & 

$4.75$

$\left(11\bar{2}0\right)$ \cite{coop1962}
 & ---\tabularnewline
\midrule 
density $\rho\,\left(\frac{\text{kg}}{\text{m}^{3}}\right)$ & 9170 \cite{clar1978} & 2200 \cite{roye1999} & 8580 \cite{zare2016} & 3980 \cite{burg1994} & 4507 \cite{wint2024}\tabularnewline
\midrule 
sound velocity $v\,\left(\frac{\text{nm}}{\text{ps}}\right)$ & $4.07^{*}$ & $5.35^{**}$ & 5.16 \cite{zare2016} & 11.2 \cite{burg1994} & 6.07 \cite{lide2004}\tabularnewline
\midrule 
accoustic impedance $Z\,\left(\frac{10^{6}\,\text{kg}}{\text{m}^{2}\,\text{s}}\right)$ & $36.1$ & $11.8$ & $44.3$ & $44.6$ & $27.4$\tabularnewline
\midrule 
heat capacity $C_{p}\,\left(\frac{\text{J}}{\text{kg}\,\text{K}}\right)$ & 330 \cite{germ1981} & 725 \cite{ande1992} & 260 \cite{haba2013} & 790 \cite{burg1994} & 530 \cite{wint2024}\tabularnewline
\midrule 

linear thermal\\
expansion coefficient. $\zeta\,\left(\frac{10^{-6}}{\text{K}}\right)$
 & $23.7$ \cite{zeus2019} & 0.49 \cite{hahn1972} & 6.89 \cite{haba2013} & 5.38 \cite{hahn1978} & 8.6 \cite{wint2024}\tabularnewline
\midrule 

thermal\\
conductivity $\kappa\,\left(\frac{\text{W}}{\text{m}\,\text{K}}\right)$
 & $5$ \cite{zeus2019} & 1.34 \cite{ande1992} & 53.3 \cite{haba2013} & 40 \cite{burg1994} & 21.9 \cite{wint2024}\tabularnewline
\midrule 

complex refractive index\\
$\tilde{n}$ at $800\,$nm
 & 

$1.84$
$+3.09\text{i}$ \cite{zeus2019}
 & 

$1.45$
$+0.00\text{i}$ \cite{mali1965}
 & 

$2.33$
$+3.24\text{i}$ \cite{weav1973}
 & 

$1.76$
$+0.00\text{i}$ \cite{mali1962}
 & 

$2.47$
$+2.53\text{i}$ \cite{palm2018}
\tabularnewline
\midrule 

complex refractive index\\
$\tilde{n}$ at $400\,$nm
 & 

$0.97$
$+1.76\text{i}$ \cite{zeus2019}
 & 

$1.47$
$+0.00\text{i}$ \cite{mali1965}
 & 

$2.53$
$+2.61\text{i}$ \cite{weav1973}
 & 

$1.79$
$+0.00\text{i}$ \cite{mali1962}
 & 

$2.09$
$+2.96\text{i}$ \cite{palm2018}
\tabularnewline
\bottomrule
\end{tabular}
\end{table}

\subsection{Optical penetration of pump and probe}
\begin{figure}[htbp]
\centering
\includegraphics[width = 0.75\textwidth]{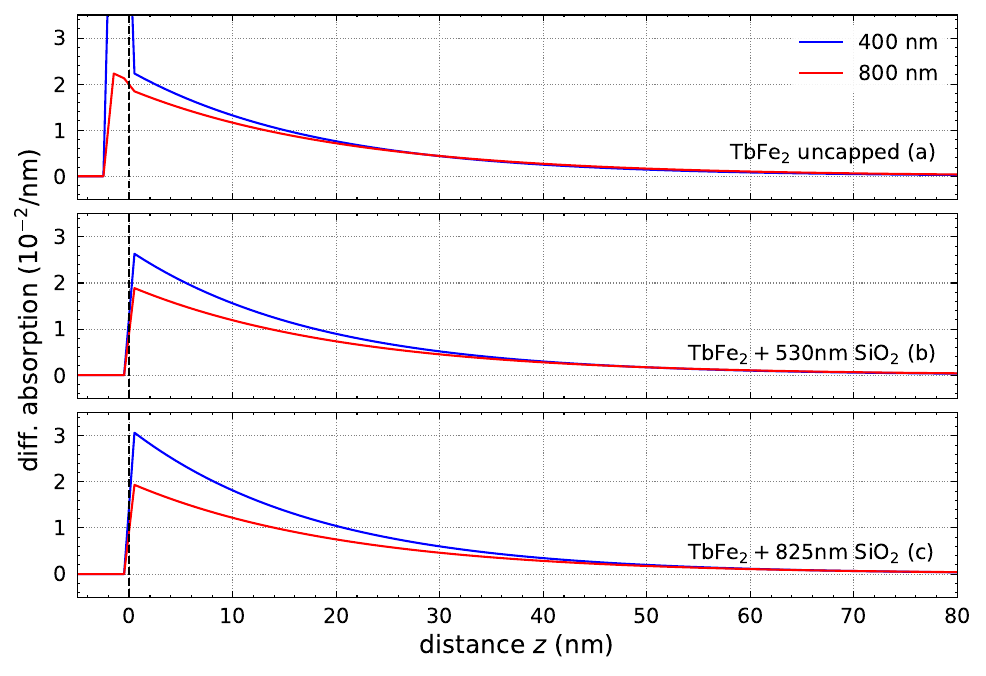}
\caption{\label{fig:abs_profiles}\textbf{Optical penetration profiles of the 800\,nm pump and 400\,nm probe light.} The optical penetration of the light into TbFe$_\mathrm{2}$ samples has been calculated via the optical transfer matrix method built into the \textsc{udkm1Dsim} python toolbox\cite{schi2021}. The depiction shows the results for all three sample configurations: uncapped (a), with a 530\,nm SiO$_\mathrm{2}$ cap (b), and with an 825\,nm SiO$_\mathrm{2}$ cap (c). The optical absorption profiles within the TbFe$_\mathrm{2}$ that starts at $z=0$\,nm are nearly identical regardless of the capping layer. \label{fig:penetration_profile}}
\end{figure}

We derived the optical excitation profiles for the p-polarized 800\,nm pump and 400\,nm probe pulses using a transfer matrix model in the \textsc{udkm1Dsim} toolbox \cite{schi2021} with refractive indices from Tab.~\ref{tab:Modeling_parameters}. As shown in Fig.~\ref{fig:penetration_profile}, the resulting profiles assume ideal flat films with atomically sharp interfaces and bulk-like properties, yet they provide a qualitative approximation of the optically probed near-surface region of the thick TbFe$_2$ films.
The calculated light penetrations depths are $18\,\text{nm}$ for the $400\,\text{nm}$ probe light, and $21\,\text{nm}$ for the $800\,\text{nm}$ pump light.

\section{LLG model details and limitations}

To obtain the modeled $m_\mathrm{z}$ components displayed in Fig.~2(c) and Fig.~3(b) of the main text, we use a simplified Landau–Lifshitz–Gilbert (LLG) model. The magnetization dynamics are described by:

\begin{equation} \frac{\partial\vec{m}}{\partial t} = -\frac{\mu_{0}\gamma}{1+\alpha^2}\,\vec{m}\times\vec{H}_\mathrm{eff} - \frac{\mu_{0}\gamma\alpha}{1+\alpha^2}\,\vec{m}\times\Bigl(\vec{m}\times\vec{H}_\mathrm{eff}\Bigr), \end{equation}

where $\vec{m} = \vec{M}/M_{S}$ is the normalized magnetization vector, $\mu_{0}$ is the vacuum permeability, $\gamma$ is the gyromagnetic ratio, and $\alpha$ denotes the Gilbert damping constant. The effective magnetic field $\vec{H}_\mathrm{eff}$ includes contributions from the externally applied field ($\vec{H}_\mathrm{Z}$), a demagnetization field ($\vec{H}_\mathrm{S}$), cubic magnetocrystalline anisotropy ($\vec{H}_\mathrm{ani}$), and magnetoelastic coupling ($\vec{H}_\mathrm{me}$).

For conceptual simplicity, our model assumes a single macrospin, representing the spatially averaged magnetization within the optically probed sample region. This macrospin approximation significantly oversimplifies the treatment of ultrafast magnetization dynamics, especially during the initial picoseconds after laser excitation. Consequently, our simplified approach neglects several complex phenomena such as: the explicit ferrimagnetic coupling between the Tb and Fe sublattices, intrinsic ultrafast demagnetization processes, electron-phonon coupling typically described within a two-temperature model framework, and the temperature-dependent reduction of both magnetocrystalline anisotropy constants ($K_{1,2}$) and magnetoelastic coupling parameters ($b_{1,2}$). As a result, this simplified model does not capture the complete non-equilibrium magnetization dynamics occurring within the first 100\,ps after laser excitation. Nevertheless, it successfully describes the magnetization response to propagating picosecond strain pulses as well as the quasi-static strain conditions prevailing at later time delays, beyond the initial transient response induced by laser excitation.

\subsection{LLG Model details}
In this subsection, we provide a concise overview of the parameters used in our Landau–Lifshitz–Gilbert (LLG) simulations. The effective magnetic field $\vec{H}\mathrm{eff}$ in the LLG equation is decomposed into distinct contributions: the externally applied field ($\vec{H}_\mathrm{ext}$), the demagnetization field ($\vec{H}_\mathrm{D}$), the cubic magnetocrystalline anisotropy field ($\vec{H}_\mathrm{ani}$), and the magnetoelastic coupling field ($\vec{H}_\mathrm{me}$). Due to the (110)-orientation of the TbFe$_\mathrm{2}$ crystallographic axes as the out-of-plane direction along along the $z$ direction in the laboratory frame, transformations into the crystallographic reference frame (denoted by a tilde) are necessary for correctly describing the anisotropy and magnetoelastic interactions. Below, we explicitly define each component of $\vec{H}_\mathrm{eff}$, followed by Table~\ref{tab:magnetic_parameters}, which summarizes both the literature values and the parameters specifically extracted from our measurements on uncapped and SiO$_2$-capped TbFe$_2$ samples. The notable differences observed between literature and experimentally extracted values for the anisotropy constants $K_{1,2}$ and magnetoelastic coupling parameters $b_{1,2}$ may originate from temperature-dependent parameter reductions\cite{engd2000} following laser excitation, as well as from structural imperfections and reduced crystallinity typically encountered in thin-film samples\cite{moug1999}.

{
\begin{subequations}
\begin{align}
\vec{H}_\mathrm{eff} &= \vec{H}_\mathrm{ext} +\vec{H}_\mathrm{D} +\vec{H}_\mathrm{ani} + \vec{H}_\mathrm{me}\,\,\,\,  \,\,\,\, \text{with} \\
\vec{H}_\mathrm{D} &= -M_{S}\,m_{z}\,\hat{z},\\[1ex]
\vec{\tilde{H}}_{\text{ani}} & =-\frac{2K_{1}}{\mu_{0}M_{S}}\left(\begin{array}{c}
\tilde{m}_{x}\left(\tilde{m}_{y}^{2}+\tilde{m}_{z}^{2}\right)\\
\tilde{m}_{y}\left(\tilde{m}_{x}^{2}+\tilde{m}_{z}^{2}\right)\\
\tilde{m}_{z}\left(\tilde{m}_{x}^{2}+\tilde{m}_{y}^{2}\right)
\end{array}\right)-\frac{2K_{2}}{\mu_{0}M_{S}}\left(\begin{array}{c}
\tilde{m}_{x}\tilde{m}_{y}^{2}\tilde{m}_{z}^{2}\\
\tilde{m}_{y}\tilde{m}_{x}^{2}\tilde{m}_{z}^{2}\\
\tilde{m}_{z}\tilde{m}_{x}^{2}\tilde{m}_{y}^{2}
\end{array}\right)\\
\vec{\tilde{H}}_{\text{me}} & =-\frac{2b_{1}}{\mu_{0}M_{S}}\left(\begin{array}{c}
\tilde{\eta}_{xx}\tilde{m}_{x}\\
\tilde{\eta}_{yy}\tilde{m}_{y}\\
\tilde{\eta}_{zz}\tilde{m}_{z}
\end{array}\right)-\frac{b_{2}}{\mu_{0}M_{S}}\left(\begin{array}{c}
\tilde{\eta}_{xy}\tilde{m}_{y}+\tilde{\eta}_{xz}\tilde{m}_{z}\\
\tilde{\eta}_{xy}\tilde{m}_{x}+\tilde{\eta}_{yz}\tilde{m}_{z}\\
\tilde{\eta}_{xz}\tilde{m}_{x}+\tilde{\eta}_{yz}\tilde{m}_{y}
\end{array}\right)
\end{align}
\end{subequations}
}
Herein, $K_{1,2}$ denote the cubic magnetocrystalline anisotropy constants and $b_{1,2}$ the magneto-elastic coupling parameters. Owing to the (110)-orientation of the TbFe$_2$ unit cells, the anisotropy and magneto-elastic fields need to be transformed into the crystallographic reference frame, denoted by a tilde. Numerical values used in our modeling are provided in Table~\ref{tab:magnetic_parameters}, along with literature values where available.

\begin{table}
%\caption{Magnetic parameters used in the Landau–Lifshitz–Gilbert simulations discussed in the main text. Literature values are shown alongside those extracted for uncapped and SiO$_\mathrm{2}$ capped TbFe$_\mathrm{2}$ samples. The low $K_{1,2}$ and $b_{1,2}$ values might be explained by their strong temperature dependence, especially following the laser excitation.\label{tab:magnetic_parameters} }
\caption{ Magnetic parameters used in the Landau–Lifshitz–Gilbert simulations discussed in the main text. Literature values are shown alongside those extracted for uncapped and SiO$_\mathrm{2}$ capped TbFe$_\mathrm{2}$ samples. The low $K_{1,2}$ and $b_{1,2}$ values might be explained by their strong temperature dependence, especially following the laser excitation, as well as an inhomogeneous crystalline structure. Differences between the individual samples and the excitation fluences might explain the different $K_{1,2}$ and $b_{1,2}$ used for modeling. \label{tab:magnetic_parameters} }
\centering{}%
\begin{tabular}{lccc}
\toprule 
 & literature & uncapped & capped\tabularnewline
\midrule
\midrule 
saturation magnetization $M_{S}\,\left(\frac{\text{MA}}{\text{m}}\right)$  & \multicolumn{3}{c}{$0.80$ \cite{clar1978}}\tabularnewline
\midrule 
cubic anisotropy const. $K_{1}\,\left(\frac{\text{MJ}}{\text{m}^{3}}\right)$ & $-12.70$ \cite{moug2000} & $-1.06$ & $-1.55$\tabularnewline
\midrule 
cubic anisotropy const. $K_{2}\,\left(\frac{\text{MJ}}{\text{m}^{3}}\right)$ & $2.08$ \cite{moug2000} & $0.17$ & $0.25$\tabularnewline
\midrule 
magnetoelastic const. $b_{1}\,\left(\frac{\text{MJ}}{\text{m}^{3}}\right)$ & $-34$ \cite{parp2021} & $-2.8$ & $-4.2$\tabularnewline
\midrule 
magnetoelastic const. $b_{2}\,\left(\frac{\text{MJ}}{\text{m}^{3}}\right)$ & $-360$ \cite{parp2021}& $-30$ & $-44$\tabularnewline
\midrule 
damping constant $\alpha$ & --- & \multicolumn{2}{c}{$2.0$}\tabularnewline
\midrule 
external magnetic field $\vec{H}_{\text{ext}}\,\left(\text{T}\right)$ & --- & \multicolumn{2}{c}{$1.3$}\tabularnewline
\bottomrule
\end{tabular}
\end{table}

\subsection{Transformation between the laboratory and the crystallographic frame of reference}
\begin{figure}
    \centering
    \includegraphics[width=0.5\linewidth]{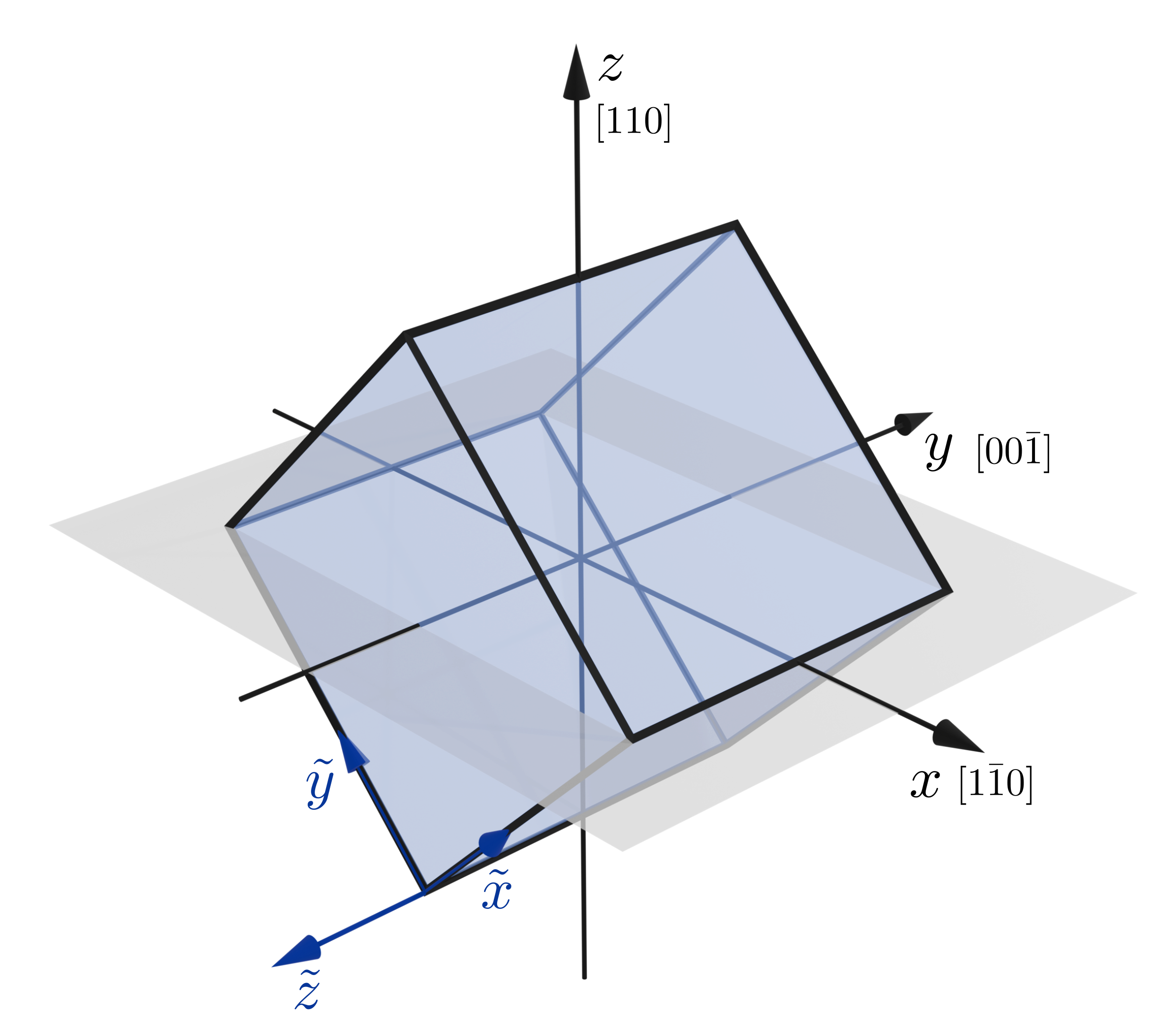}
    \caption{\textbf{Geometric representation of the sample and crystallographic coordinate systems:} The sample axes (black) are defined with the $z$–axis normal to the thin film sample surface that is indicated by the light-gray plane. Therefore, $z\parallel[110]$, and the projections of the sample $x$– and $y$–axes coincide with the $[1\bar{1}0]$ and $[00\bar{1}]$ directions, respectively.
    The crystallographic axes (blue) are aligned with the main cubic axes $\langle 100 \rangle$.}
    \label{fig:geometry}
\end{figure}

As briefly mentioned in the main text, the expressions for the magnetocrystalline anisotropy field ($\vec{\tilde{H}}_{\mathrm{ani}}$) and the magnetoelastic field ($\vec{\tilde{H}}_{\mathrm{me}}$) are naturally defined within the crystallographic reference frame, where coordinate axes align with the principal cubic crystal axes. However, the LLG simulations are conducted in the laboratory frame, where the TbFe$_2$ films are oriented such that their crystallographic (110)-axis coincides with the laboratory $z$-axis that is defined to be perpendicular to the film surface. The two coordinate systems and their orientation with respect to the unit cell and sample surface are depicted in Fig.~\ref{fig:geometry}.

To correctly incorporate all effective magnetic field components, it is necessary to perform a coordinate transformation between these two reference frames. This transformation involves a composite rotation consisting of a clockwise rotation around the laboratory $x$-axis by $90^\circ$, followed by a clockwise rotation around the laboratory $z$-axis by $45^\circ$. Expressed in matrix form, this combined rotation is given by:

\begin{align}
R & =\underset{R_{z}\left(-45{^\circ}\right)}{\underbrace{\left(\begin{array}{ccc}
\cos\left(-45{^\circ}\right) & -\sin\left(-45{^\circ}\right) & 0\\
\sin\left(-45{^\circ}\right) & \cos\left(-45{^\circ}\right) & 0\\
0 & 0 & 1
\end{array}\right)}}\underset{R_{x}\left(-90{^\circ}\right)}{\underbrace{\left(\begin{array}{ccc}
1 & 0 & 0\\
0 & \cos\left(-90{^\circ}\right) & -\sin\left(-90{^\circ}\right)\\
0 & \sin\left(-90{^\circ}\right) & \cos\left(-90{^\circ}\right)
\end{array}\right)}}\\
 & =\left(\begin{array}{ccc}
\frac{1}{\sqrt{2}} & 0 & \frac{1}{\sqrt{2}}\\
-\frac{1}{\sqrt{2}} & 0 & \frac{1}{\sqrt{2}}\\
0 & -1 & 0
\end{array}\right)
\end{align}

Using this rotation matrix, the magnetization vector transforms from the laboratory frame ($\vec{m}$) into the crystallographic frame ($\vec{\tilde{m}}$) according to:

\begin{equation}
\vec{\tilde{m}}=R\vec{m}=\frac{1}{\sqrt{2}}\left(\begin{array}{c}
m_{x}+m_{z}\\
m_{z}-m_{x}\\
-m_{y}
\end{array}\right).
\end{equation}

The resulting anisotropy field $\vec{\tilde{H}}_{\text{ani}}$ must
then be transformed back into the sample frame $\vec{H}_{\text{ani}}\left(\vec{m}\right)=R^{T}\vec{\tilde{H}}_{\text{ani}}\left(\vec{\tilde{m}}\right)$
by the inverse rotation matrix $R^{-1}{=}R^{T}$.

For the magnetoelastic field, the same transformation is applied, and in addition, the strain tensor must be transformed by $\tilde{\eta}=R\eta R^{T}$. Notably, a longitudinal strain along $z$ in the sample frame partially corresponds to a shear strain, i.e. non-zero off-diagonals in the crystallographic frame:
\begin{equation}
\tilde{\eta}=R\left(\begin{array}{ccc}
0 & 0 & 0\\
0 & 0 & 0\\
0 & 0 & \eta_{zz}
\end{array}\right)R^{T}=\frac{1}{2}\left(\begin{array}{ccc}
\eta_{zz} & \eta_{zz} & 0\\
\eta_{zz} & \eta_{zz} & 0\\
0 & 0 & 0
\end{array}\right).
\end{equation}
This transformation illustrates how longitudinal strain in the laboratory frame produces significant shear components in the crystallographic reference frame, thus activating the $b_{2}$ magnetoelastic coupling  parameter that is approximately ten times larger than $b_\mathrm{1}$. Consequently, in the laboratory frame, the magnetoelastic field acquires not only the expected $z$-component but also a nonzero $x$-component:
\begin{equation}
\vec{H}_{\text{me}}\approx-\frac{1}{\sqrt{2}}b_{2}\eta_{zz}\left(\begin{array}{c}
-m_{x}\\
0\\
m_{z}
\end{array}\right).
\end{equation}

\section{Supplementary data}
\begin{figure}[htbp]
\centering
\includegraphics[width = \textwidth]{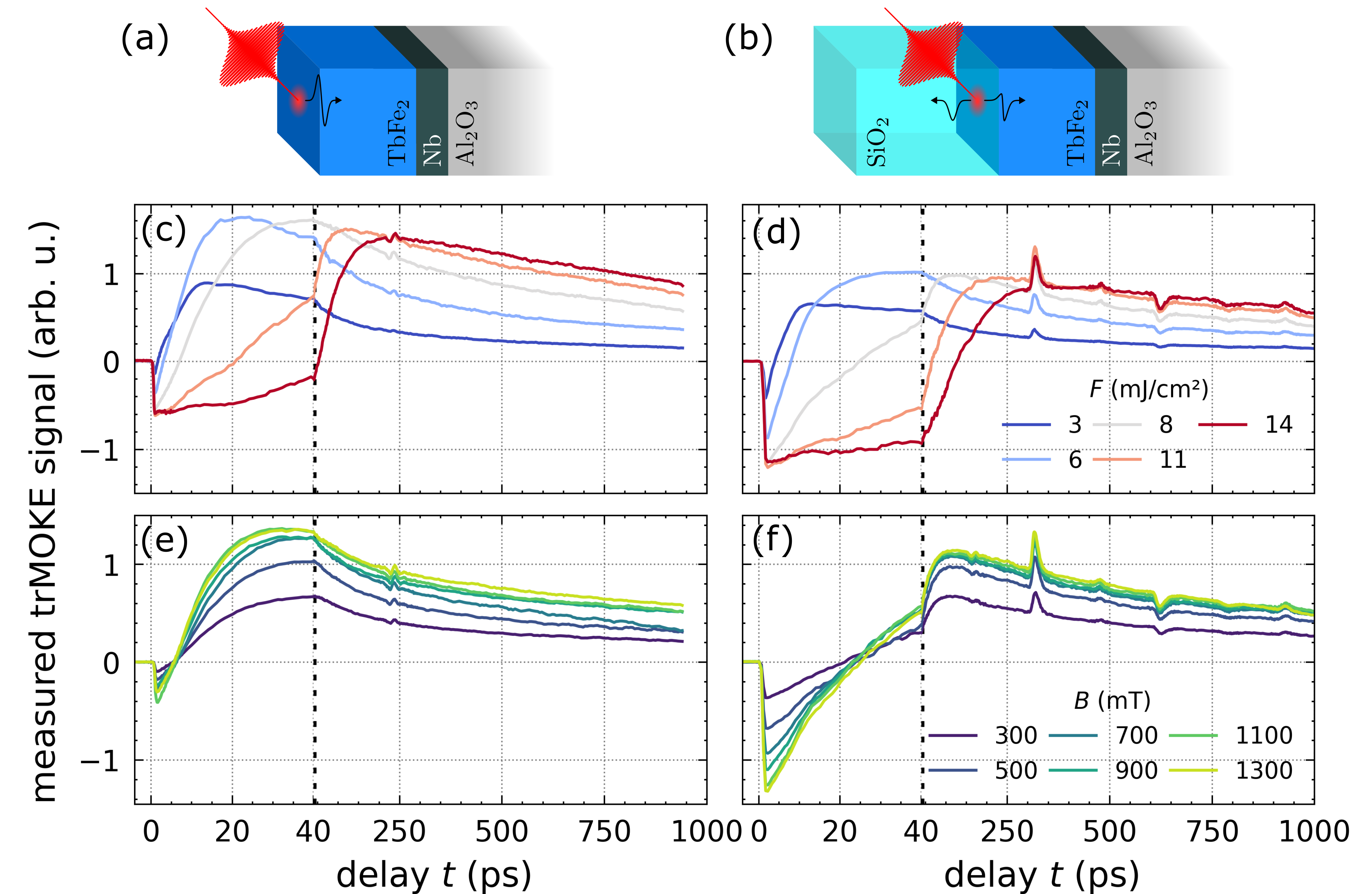}
\caption{\label{fig:fluence_series}\textbf{Fluence and field dependent trMOKE data:} (a, b) Schematic cross sections of the uncapped (left) and SiO$_\mathrm{2}$-capped (right) TbFe$_\mathrm{2}$ sample geometries. (c, d) Time‐resolved magneto‐optical Kerr effect (trMOKE) signals measured at different laser excitation fluences (3-14$\,\mathrm{mJ/cm^{2}}$). (e, f) trMOKE signals at 8\,$\mathrm{\frac{mJ}{cm^2}}$ under varying external magnetic fields (300–1100\,mT). Note the axis break indicated by the indicated by the vertical dashed line, which allows to display the trMOKE response at early times combined with the late delays. The signatures or the picosecond strain pulses within the trMOKE signal that are discussed in the main text are present regardless of the exact experimental condition.}
\end{figure}

In the main text, we present the representative trMOKE response for an excitation fluence of 8\,mJ/cm$^2$ at the maximum available B-field of approximately 1300\,mT. For completeness, Fig.~\ref{fig:fluence_series} displays additional field-dependent and fluence-dependent measurement results for both uncapped and capped Terfenol samples. Although the initial trMOKE response within the first 40\,ps is strongly fluence dependent, the characteristic modulations triggered by picosecond strain pulses that are discussed in the main manuscript remain consistent for each fluence. The main effect of the external magnetic field is that it primarily scales the signal amplitude without qualitatively altering the trMOKE response. These findings are in good agreement with the previous works by Tymur Parpiev and co-workers.\cite{parp2017b, parp2021}

\begin{figure}[htbp]
\centering
\includegraphics[width = 1\textwidth]{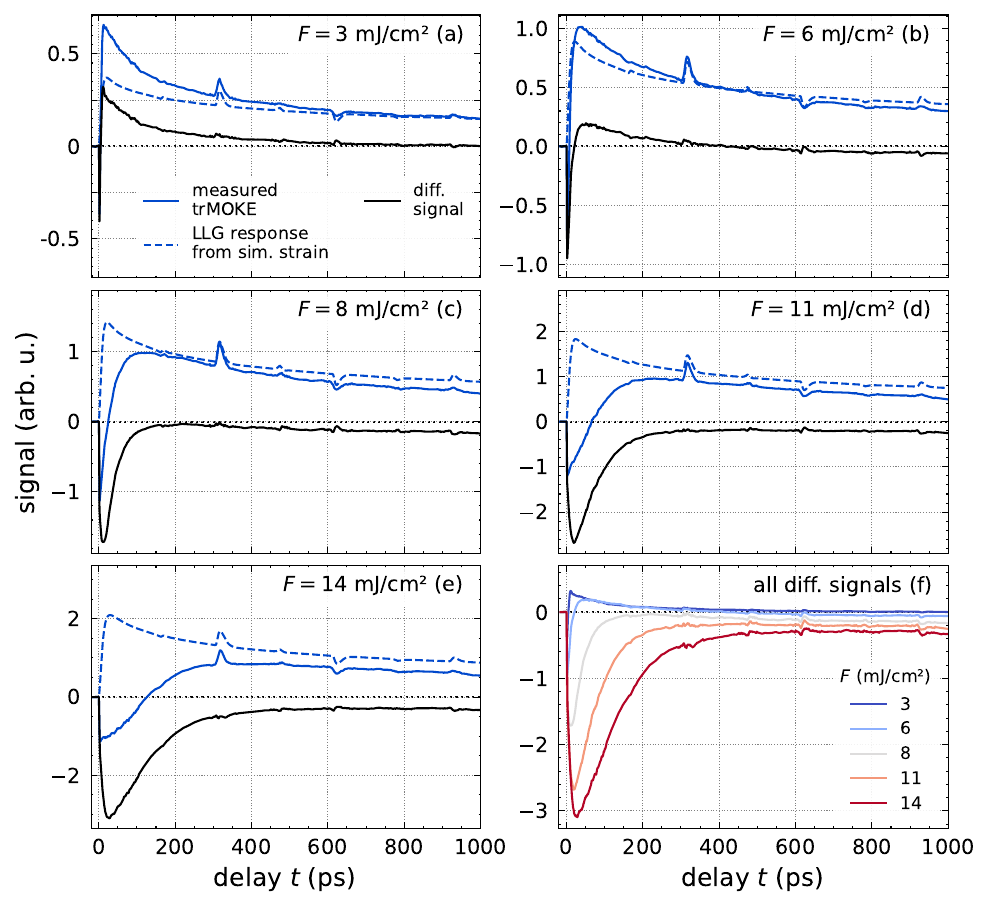}
\caption{\label{fig:fluence_separation}\textbf{Analysis of the trMOKE response in glass-capped $\mathrm{TbFe_2}$ for different fluences.} This figures shows the fluence-dependent separation of the trMOKE signal into the magneto-elastically driven contribution and the intrinsic magnetization response via demagnetization and anisotropy changes for the 825nm SiO$_2$ capped Terfenol structure. The procedure follows the steps exaplained in the main text (Fig. 3).\\
Panels (a-e) present the trMOKE signals, modeled magnetization responses and corresponding difference signals for varying fluences. Panel (f) combines all extracted difference signals where the magneto-elastic contribution has been subtracted for a direct comparison.
}
\end{figure}

 Although a full discussion of the complex trMOKE evolution in Terfenol is beyond the scope of this work, we examplarily demonstrate that the magneto-elastic contribution can subtracted for all measurements in the excitation fluence series for the glass-capped sample structure. Using the approach described and illustrated by Fig.~3 in the main text, the residual magnetization response is obtained as shown in Fig.~\ref{fig:fluence_separation}. Notably, the extracted signal changes sign at low fluences, likely due to an interplay between sublattice-dependent demagnetization, which reduces the magnetization amplitude, and a reduction of the temperature-dependent magnetocrystalline anisotropy that induces an increased out-of-plane tilt and thus a larger $m_\mathrm{z}$. At larger fluences $\gtrsim 8\,\mathrm{\frac{mJ}{cm^2}}$, demagnetization likely becomes dominant, and the tilting is not sufficient to raise $m_z$ above the initial level.
The speed of the signal recovery after the initial demagnetization decreases significantly with fluence, indicating slower dynamics as the overall magnetization is reduced. We attribute the fluence-dependent offset at long delays to our LLG model's omission of laser-induced demagnetization, which would shorten the effective magnetization vector. A definitive and complete interpretation of this signal requires further measurements, ideally with sublattice-sensitive probing or in a saturation field ($\mu_0H_{\text{ext}}>2.1$\,T) that fully aligns the magnetization out-of-plane in order to unequivocally extract the laser-induced demagnetization response.

\clearpage

%\bibliography{references.bib}

\end{document}